\documentclass[review]{elsarticle}
\usepackage{lineno}
\modulolinenumbers[5]

\usepackage{amsmath,amsfonts,breqn}
\usepackage{bbold}
\usepackage{stmaryrd}
\usepackage{textcomp}
\usepackage{graphicx}
\usepackage{mathtools}
\usepackage{multicol}
\usepackage{subfig, wrapfig}
\usepackage{epstopdf}
\usepackage{relsize}
\usepackage{wasysym}
\usepackage{multirow}
\usepackage{rotating}
\usepackage{dashrule}
\usepackage{xcolor}
\usepackage{comment}
\usepackage[top=1in, bottom=1in, left=1in, right=1in]{geometry}

\usepackage[colorlinks,citecolor=red,urlcolor=blue,bookmarks=false,hypertexnames=true]{hyperref}

\title{Deriving a representative variant for the functional safety development according to ISO 26262}

\author[]{Felix~S.~Schranner\corref{cor1}}
  \ead{schranner@techcos.de}
  \address{Technical University of Munich, 80333~Munich,~Germany}
   \author[]{Alireza Abassi Misheni}
  \ead{misheni@techcos.de}
  \author[]{Jork Warnecke}
  \ead{warnecke@techcos.de}
	\address{techcos GmbH, 80807~Munich,~Germany}
  \cortext[cor1]{Corresponding author. Phone: +49 (89) 4141 842- 58, e-mail: schranne@tum.de}

\journal{the Journal of Reliability Engineering \& System Safety}
\begin{document}
\begin{frontmatter}
 
\begin{abstract}
The emerging mass individualization \cite{KOREN201564} in series produced road vehicles, superseding mass customization, entails an increase in variants. The question ``{}How may the functional safety development effort, corresponding to the variant numbers, be tackled while complying to the standard ISO 26262:2018? \cite{iso26262_2nd}''{} motives this work.

The lever on containing the effort is most effective when applied in the development life-cycle as early as possible. The initiating functional safety artifact is the item definition. Strategic clustering of variants into a few or even a single item-definition(s) is feasible by exploiting platform development. Yet, the functional safety development effort for each defined item may only be limited when basing all subsequent phases on one single representative variant. 
In this work we particularly elaborate on answering the question ``{}Under what conditions can a representative variant be defined?''{}, accompanied by the notion of ``{}How is it defined?''. Therefore, we assess the ISO 26262:2018 \cite{iso26262_2nd} for deriving prerequisites on a representative variant. Furthermore, we propose a structured and efficient method to determine representative variants given vehicle parameters and a framing context. The method is exemplified for two easily comprehensible use-case situations in drivetrain development. The characteristic equations may be used for future drivetrain functional safety development.
 
\end{abstract}
\end{frontmatter}


\section{Introduction}\label{sec:Introduction}

As stated in the introduction of ISO 26262:2018 \cite{iso26262_2nd}, the increasing technological complexity, software content and mechatronics implementation in road vehicles entails an increase in risks due to systematic and random hardware failures. Yet, safety, described as the absence of unreasonable risks, has become one of the key factors for a consumer purchase decision \cite{KOPPEL2008994, SHAABAN2019493}. Thus, the automotive industry is urged to ensure safety. The ISO 26262:2018 \cite{iso26262_2nd} gives guidance for establishing functional safety in road vehicles. Therefore, it assumes a V-model \cite{Forsberg1991} development process and standardizes thereupon the safety lifecycle process of automotive manufacturers \cite{LUO2016151}. 
The second edition of the ISO 26262, \cite{iso26262_2nd} generalizes the management and engineering activities, principles and techniques for passenger vehicles formulated in ISO 26262 first edition \cite{iso26262_1st} to road vehicles with a maximum gross weight of 3500 kg. See \cite{debouk2018} for an overview of major changes between ISO 26262:2011, i.e. first edition, \cite{iso26262_1st} and ISO 26262:2018, i.e. second edition, \cite{iso26262_2nd}, or e.g. \cite{KAFKA20122} for a review of ISO 26262:2011 \cite{iso26262_1st}.
 
Let alone the increasing complexity of software-intensive safety-critical embedded systems \cite{LUO2016151}, the increase in complexity attributes significantly to the growth of variants of vehicle derivates \cite{Graf2015}.
In modular development of series produced road vehicles multiple variants derive from the same platforms to fulfill specific customer needs. For a review of product design paradigm refer to \cite{KOREN201564}. Mass customization in series produced vehicles considers variable hardware, e.g. lights, drivetrain components, the wheelbase, the ground clearance, behavior, e.g. parameterisations or different operational modes, or even entire functions. 

Assuring safety, i.e. compliance to \cite{iso26262_2nd}, for road vehicles with nowadays large variance and complex embedded systems is costly and time-consuming \cite{Drabble2009}. The business  and program management, however, are interested in achieving a high compliance coverage while keeping costs low \cite{LUO2016151}.
Multiple variants may be clustered in a single item definition on the basis of common requirements, see \cite{iso26262_2nd} Part 3 Section 5.
To limit development effort, the subsequent phases of functional safety development according to ISO 26262:2018 \cite{iso26262_2nd} have to be pursued with only a few representative variants of the cluster. Yet, how and on what basis is a representative variant defined?

Yoe \cite{Yoe2019}, for example, discusses representative scenarios commonly employed in risk analysis. Among the failure-scenarios are the best-case, most likely and worst-case scenario. Thereby, the worst-case scenario is the one best-known \cite{Yoe2019}.
Design of experiment (DoE) targets towards minimizing the number of test candidates while concurrently acquiring most information by appropriately combining input factors \cite{Carlson2001, C.F.JeffWu2009}. In the context of validation testing, a worst-case parameter combination permits to draw the conclusion if or if not a device is usable, see e.g. IEC 62366:2007 \cite{IEC62366:2007} Annex D or \cite{Loring2018}, a process at the limit of acceptable operating conditions yields a product as specified, see e.g. \cite{Gardner1998}, or a system performs within the specified bounds, see \cite{Fitzpatrick2018}. 

This publication elaborates in the first part on the considerations to be made to determine a representative variant that is in accordance to \cite{iso26262_2nd}, see Section \ref{sec:ASIL_ISO26262}. Based thereupon, an efficient method is formulated that agrees with \cite{iso26262_2nd} in Section \ref{sec:WC_ISO26262}. In the second part of this work, basic models for selecting the representative variant for functional safety development for the drivetrain are derived, see Section \ref{sec:DriveTrain-WC}.
For an application example, a set of variants are clustered as one item. Their use-case dependent representative variants are determined in Section \ref{Sec:application_example}. Concluding remarks are stated in Section \ref{sec:Conclusion}.

\section{A representative variant according to ISO 26262}\label{sec:Rep_Vehicle_basics}
\subsection{Basic considerations to derive a representative variant}\label{sec:ASIL_ISO26262}
According to hazard theory, see e.g. \cite{CliftonA.Ericson2005}, a hazardous event consists of the three necessary elements: a hazard source, an (initiating) mechanism and a target. In the context of \cite{iso26262_2nd}, the hazard source is the item described in the item definition at vehicle level according to ISO 26262:2018 \cite{iso26262_2nd}-3 Sec. 5. The initiating mechanism is the malfunction of the item in the context of a certain operational mode or situation, stating a use-case. The target is the potentially harmed human(s).

To define a hazardous event, the item without internal safety mechanisms is evaluated with the hazard analysis and risk assessment (HARA), described in ISO 26262:2018 \cite{iso26262_2nd} Part 3 Sec. 6.
According to \cite{iso26262_2nd} Part 3 Sec. 6.4.3, hazards are classified conservatively with respect to severity (S), probability of exposure (E) and controllability (C), ``{}i.e. whenever there is a reasonable doubt, a higher S, E or C classification is to be chosen.''{} \cite{iso26262_2nd}. Thereby, \cite{iso26262_2nd} circumvents the shortfall of not providing a truly objective rating methodology, yet, instead leaving the ASIL classification to the skills and the mental model of domain technical experts \cite{KHASTGIR2017166}. 
Moreover, the``{} level of rigor required for higher ASIL values is considerably high as compared to a lower ASIL value. Therefore, the automotive industry is always driven towards lower
ASIL values in order to keep their development costs down. This inherent bias can also sometimes lead to an inconsistency in the ASIL ratings.''{}\cite{KHASTGIR2017166}.\\
Nevertheless, to determine S, the potential harm caused to each person at risk is evaluated, see \cite{iso26262_2nd} Part 3 Sec. 6.4.3.2. If applicable, the potential injuries shall be clustered to derive the highest S-classification. A structured aid to determine S is the Abbreviated Injury Scale (AIS) \cite{BAKER1974} provided e.g. in Appendix B of \cite{iso26262_2nd} Part 3. Alternative ratings may be found e.g. in \cite{IEC61508-2nd} Table E.1. Also, to classify S \cite{SAEInternational2018} tabulates data of vehicle velocities and collision types to severity. Yet, the parameters are not sufficiently exhaustive \cite{KHASTGIR2017166} to objectively determine S. 
According to \cite{iso26262_2nd} Part 3 Sec 6.4.2.1 ``{}The operational situations and operating modes in which an item's malfunctioning behavior will result in a hazardous event shall be described; ...''{}.\\
Determining E thus requires analysis of e.g. driving scenarios and conditions. The German \textit{Verband Der Automobilindustrie e.V.} (VDA) has gathered statistical data and derived the \textit{Situationskatalog E-Parameter nach ISO 26262-3} \cite{Sitkat-2015} for representative use-case scenarios in conjunction with E-classifications.\\
The controllability (C) is classified based on the probability that a representative driver is able to retain or regain control of the vehicle if a hazardous event occurs, or other road users in the vicinity are able to influence the situation such to avoid the mishap, see \cite{iso26262_2nd} Part 3 Sec 6.4.3.8 and Annex B.4. The road user or driver ability to act is the key determining factor, i.e. human factors given technical means, see e.g. \cite{ROSEN201125, Schaap2008, Young2007}. In practice the controllability may be evaluated by expert judgment of well-experienced testers accompanied with statistical data.
However, even on basis of well-tabulated data, classification is prone to inter-rateability and intra-rateability variations \cite{ERGAI2016393}, see \cite{KHASTGIR2017166}, further justifying a pessimistic classification.

The situations and modes are identical for the unity of the variants of one item given the same requirements. Thus, E is independent from technical realizations. Yet, the severity S and the controllability C both depend on the technical realization of the item, see e.g. \cite{KHASTGIR2017166, Johansson2016, McGehee2000, ROSEN201125, Young2007}. I.e. the hazardous situation needs to be classified for the ASIL with each of the variants of the item definition.

However, \cite{iso26262_2nd} Part 3 Sec. 6.4.4.1 states that ``{}A safety goal shall be determined for each hazardous event with an ASIL evaluated in the hazard analysis and risk assessment. If similar safety goals are determined, these may be combined into one safety goal.''{} The amendment ``{}The ASIL determined for the hazardous event shall be assigned to the corresponding safety goal. If similar safety goals are combined into a single one, in accordance with 6.4.4.1, the highest ASIL shall be assigned to the combined safety goal.''{}, \cite{iso26262_2nd} Part 3 Sec. 6.4.4.2,  explicitly encourages clustering. Consequently, the clustering of similar hazardous events is compliant to \cite{iso26262_2nd}. 
Thereby, the worst-case hazardous event, i.e. the one with the highest ASIL-classification, is representative for the cluster.

Again, two of the three hazard elements (hazard source, initiating mechanism, and target) are vehicle-independent: the use-case and target. 
Given these two hazard elements, the variant with the biggest hazardous potential completes the worst-case event. 
On basis of these three elements, a representative ASIL-classification is assessed. 

\subsection{Efficiently constructing a representative variant}\label{sec:WC_ISO26262}
How can the worst-case variant be determined efficiently? This implies that performing a HARA for each variant needs to be unnecessary; instead, one HARA on basis of the worst-case variant shall suffice.

Constructing a worst-case scenario relies on conservatism to credit for uncertainty in input parameters and the analyst’s judgment to choose that set of input values that yields the worst possible outcome from a model, see \cite{Yoe2019}. According to Gardner \textit{et al.} \cite{Gardner1998} ``{}The worst-case approach [...] allows examination of all of the critical [...] variables together, thus ensuring that additive effects and interactions are tested for.''{}
Loring \cite{Loring2018} analyzes the definitions of \textit{worst-case scenario} in the context of usability testing for medical devices based on IEC 62366:2007 \cite{IEC62366:2007} Annex D and the \textit{Guidance Document Applying Human Factors and Usability Engineering to Medical Devices} \cite{FDA2016}. She summarizes that for both, risk assessment and validation, parameters with the highest potential for harm, in this work's context highest ASIL, and those of actual use ought to be simulated or chosen.  

Actual use-case situations are described in the agreed-upon situation catalogue \textit{Situationskatalog E-Parameter nach ISO 26262-3}\cite{Sitkat-2015}. Applying \cite{Sitkat-2015} leads to worst-case environmental parameters. 
According to ISO 26262:2018 \cite{iso26262_2nd} Part 3 Sec. 6.4.3.8 NOTE 3 cases of reasonably foreseeable misuse are also to be considered. Thereby, cases of reasonably foreseeable misuse need to be derived from actual use-case situations such that the parameters result from a boundary value analysis. The example of ``{}not keeping the required distance to the vehicle in front as a common behavior''{} is provided in \cite{iso26262_2nd} Part 3 Sec. 6.4.3.8 NOTE 3. It derives from the actual use-case of two vehicles following another. Moreover, this case is an observed \textit{common behavior}.
Note that these parameters are variant-independent, yet, important for constructing the worst-case scenarios.

\begin{figure}
\centering
\includegraphics[width=0.9\linewidth]{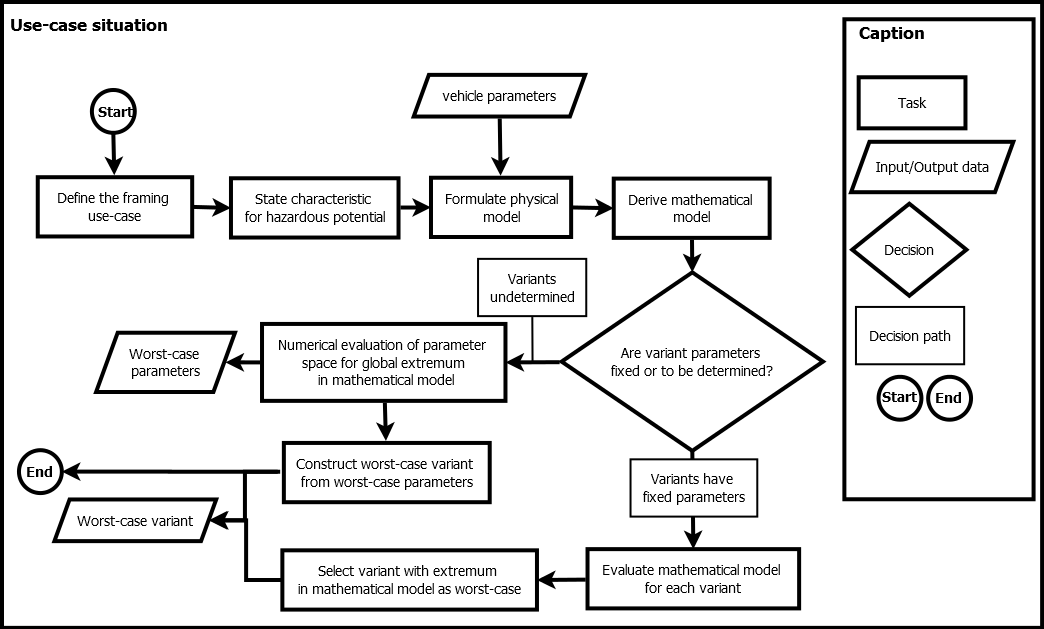}
\caption{Process to determine representative worst-case variant in use-case situation from use-case, safety-critical characteristic and vehicle parameters.}
\label{Fig:Worst-Case-Derivate-Selection}
\end{figure}

Figure \ref{Fig:Worst-Case-Derivate-Selection} depicts the steps for determining the representative variant. 
As a first step, the framing use-case is defined, i.e. the operational mode, driving scenario and conditions in which a fault occurs. 
Next, the characteristic that determines the biggest hazardous potential in a given use-case is formulated. 
Based on the analyst’s judgment \cite{Yoe2019}, a physical model is constituted. Therefore, contextual assumptions are drawn to filter key influential parameters.
Next, the to be assessed characteristic is modeled mathematically. 

Assume the general case that multiple vehicle-parameters characterize the variants. Moreover, each parameter is independent and has a certain range of validity.
Even with simplifying assumptions, the parameters may define a high-dimensional, nonlinear mathematical model. 
For efficient approaches to determine the global worst-case parameter combination refer to \cite{RinnooyKan1987, Moles2003, Jones1998, neal1997monte, Josef2017, Schranner2016}.
With the worst-case parameter combination the worst-case representative variant is stated.\\
Alternatively, the item definition may detail the parametric realization of specific variants. Consequently, the mathematical model can be evaluated directly for each variant.
A direct comparison of the characteristic value for the variants leads to the one with the biggest hazardous potential.

\subsection{Consequence for subsequent phases}\label{sec:Consquences}
With the worst-case representative variant at hand, the hazardous situations are evaluated, their ASIL derived and the safety goals formulated, see \cite{iso26262_2nd}, Part 3, Sec. 6. According to \cite{iso26262_2nd}, Part 3, Sec. 7.2 the safety measures developed within the scope of the functional safety concept need to comply with the safety goals. Hence, the safety measures are developed according to the representative variant. Thereby, these are stated comprehensively. In accordance with the functional safety concept, the product development at system level is pursued on basis of the worst-case representative variant. The safety analysis as part of phase \cite{iso26262_2nd} Part 3 Sec. 6.4 may be used to detail variant-specifics. 

The integration and testing phase, see \cite{iso26262_2nd} Part 4 Sec. 7 is variant-independent. A worst-case approach does not guarantee completeness of verification.
 
The safety validation is the final development phase at system level. It provides the evidence that the safety goals are achieved at the vehicle level by means of the integrated safety measures.  Validation of safety goals is applied to the item integrated at the vehicle level and the validation plan includes test procedures for each safety goal with a pass/fail criterion. Thereby, the evidence of appropriateness of the functional safety concept is provided. 
\cite{iso26262_2nd} Part 4 Sec. 8 stresses that the representative context and representative variant suffice for validation, see particularly \cite{iso26262_2nd} Part 4 Sec. 8.4.1 and 8.4.2.

\section{Representative variant for functional safety drivetrain development}\label{sec:DriveTrain-WC}
Within the following, models characterizing a vehicle's hazardous potential in the domain of drivetrain development are derived. We specifically study the use-cases ``{}drivetrain fault entailing an unintended acceleration with, without loss of traction''{}, respectively. 
In conjunction with vehicle parameters these use-cases state sufficiently complex scenarios. On the basis of well-established assumptions, the characteristics are formulated with key vehicle parameters. These are in agreement with the findings of \cite{Johansson2016, McGehee2000, ROSEN201125}.

Note that the process for use-case situation assessment of a fault unintentionally decelerating a road vehicle is as described within the following. One merely needs to replace the accelerating with the decelerating torque and the applicable transmission ratios.

\subsection{Use-Case ``{}Drivetrain fault entailing an unintended acceleration with loss of traction''{}}\label{sec:tractionloss_model}
\subsubsection{The framing use-case scenario assumptions}\label{sec:use-case-assumptions-lossoftraction}

The scenario of driving a $90^{\circ}$-turn on an inner city road with low friction constitutes the underlying worst-case driving situation. The latter condition is listed as $FS020$ \textit{Driving on road with low friction}. Loss of traction is facilitated due to a friction coefficient of $\mu < 0.5 +/- 0.1$. The situation is evaluated as $E3$, see \cite{Sitkat-2015}. For this analysis, a friction coefficient of $\mu = 0.4$ is assumed. 
A $90^{\circ}$-turn minimizes the curve radius to approximately $R = 4~m$. The road bank angle is negligible for the considered inner-city road. 
In \cite{Richter2016}, Chapter 8.5.2 the minimum curve radius is correlated to the driving velocity. One finds that a velocity of approximately $V = 12.12~kph$ correlates to a curve with a radius of $R = 4~m $. 
The drivetrain faults to allow the maximum engine torque. The acceleration is large enough to overcome traction. Moreover, the mass of the variant is the curb weight.

\subsubsection{Characteristic of hazardous potential}
According to friction circle theory, see e.g. \cite{Schramm2014,Rajamani2012}, the exchangeable forces between a tire and the road in the contact patch are bound by the friction force $F_{\mu}$:  
\begin{equation}
F_{\perp}^{2}+F_{\parallel}^{2} \leq F_{\mu}^{2}. 
\label{eq:Kamm_Circle}
\end{equation}
$F_{\parallel}$ and $F_{\perp}$ are the accelerating (longitudinal) forces and the cornering (transversal) forces, respectively.

Traction is overcome if $\frac{F_{\perp}^{2}+F_{\parallel}^{2}}{F_{\mu}^{2}} > 1$. 
When at least one wheel pair of a vehicle loses traction, i.e. the front or rear wheels, it diverges from its original driving path. Consequently it releases its hazardous potential to road users. 
We propose 
\begin{equation}
P_{\mu}=\frac{F_{\perp}^{2}+F_{\parallel}^{2}}{F_{\mu}^{2}}, 
\label{eq:Kamm_Circle2}
\end{equation}
to characterize the hazardous potential of each variant. 

Note, the vehicle parameters that best describe the hazardous potential of a vehicle are the kinetic energy, the maximum acceleration and the friction, see \cite{Johansson2016, McGehee2000, ROSEN201125}. The kinetic energy of the vehicle-system consists of the kinetic energy of the drivetrain and the vehicle in motion. 
Only due to the tire forces, the vehicle can translate the drivetrain energy, i.e. acceleration, into controllable kinetic energy of the vehicle in motion. Also, the kinetic energy of the vehicle in motion can only be controlled by means of the tires. Hence, the controllability depends on the ability to exchange forces between the road and the vehicle to maintain an intended driving path. If the left-hand side of Eq. \ref{eq:Kamm_Circle} exceeds its right-hand side, excess hazardous kinetic energy is in the system, which is, moreover, uncontrollable.   
Thus, a high value of $P_{\mu}$ correlates to a big hazardous potential and a higher classification of S and C.

\subsubsection{Physical Model}\label{sec:physical_model}

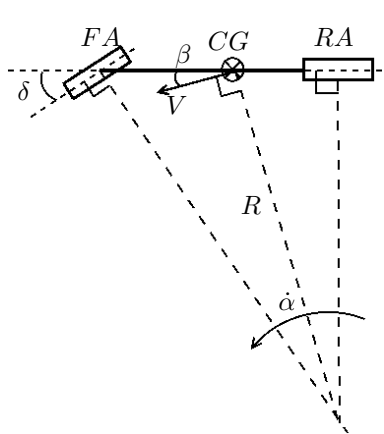
\begin{wrapfigure}[9]{l}{5cm}
\centering
\def\svgwidth{5cm}
  \makeatletter%
  \providecommand\color[2][]{%
    \errmessage{(Inkscape) Color is used for the text in Inkscape, but the package 'color.sty' is not loaded}%
    \renewcommand\color[2][]{}%
  }%
  \providecommand\transparent[1]{%
    \errmessage{(Inkscape) Transparency is used (non-zero) for the text in Inkscape, but the package 'transparent.sty' is not loaded}%
    \renewcommand\transparent[1]{}%
  }%
  \providecommand\rotatebox[2]{#2}%
	\setlength{\unitlength}{\svgwidth}%
  \global\let\svgwidth\undefined%
  \global\let\svgscale\undefined%
  \makeatother%
  \begin{picture}(1,1.04177542)%
    \setlength\tabcolsep{0pt}%
    \put(0,0){\includegraphics[width=\unitlength,page=1]{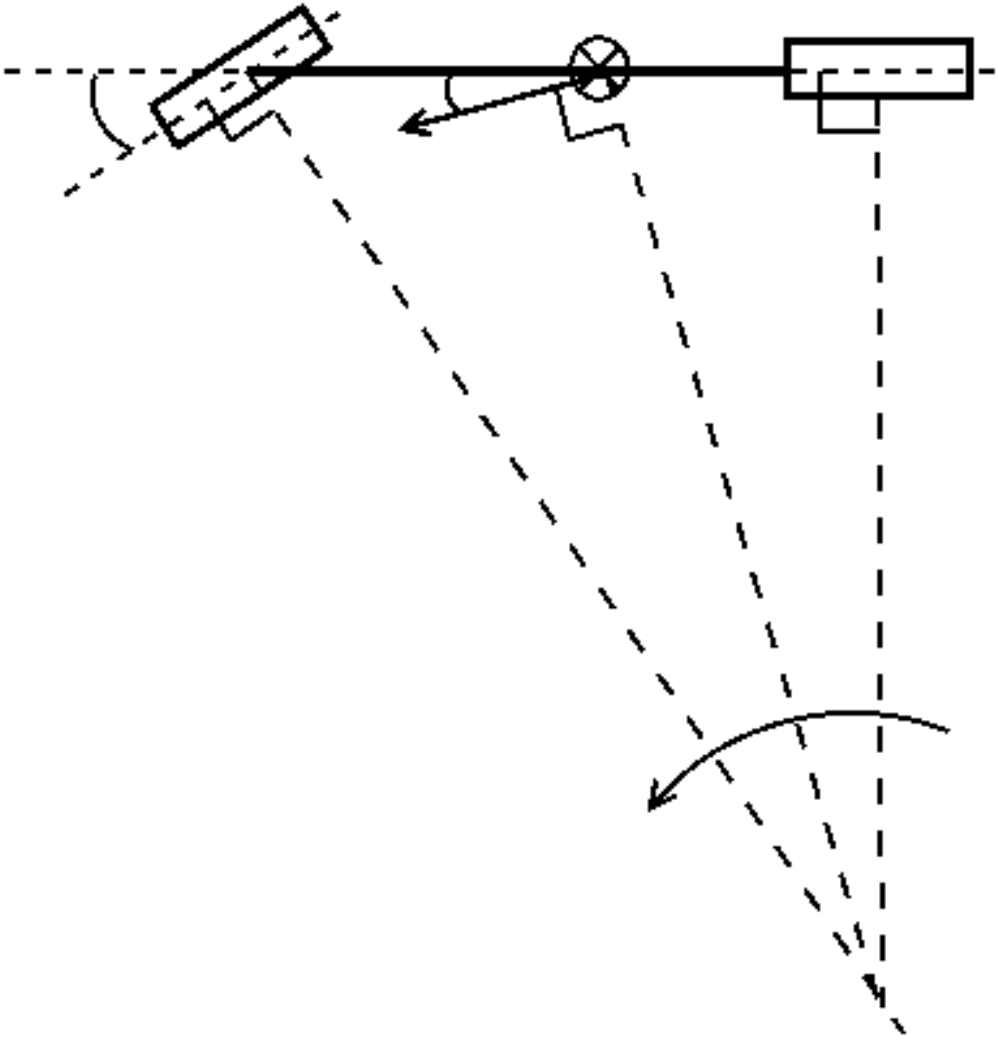}}%
    \put(0.44936395,0.9877547){\color[rgb]{0,0,0}\makebox(0,0)[lt]{\smash{\normalsize{$\beta$}}}}%
    \put(0.72439347,0.33201421){\color[rgb]{0,0,0}\makebox(0,0)[lt]{\smash{\normalsize{$\dot{\alpha}$}}}}%
    \put(0.62419784,0.58782099){\color[rgb]{0,0,0}\makebox(0,0)[lt]{\smash{\normalsize{$R$}}}}%
    \put(0.53607751,1.02319949){\color[rgb]{0,0,0}\makebox(0,0)[lt]{\smash{\normalsize{$CG$}}}}%
    \put(0.19779418,1.03299061){\color[rgb]{0,0,0}\makebox(0,0)[lt]{\smash{\normalsize{$FA$}}}}%
    \put(0.81730034,1.03494886){\color[rgb]{0,0,0}\makebox(0,0)[lt]{\smash{\normalsize{$RA$}}}}%
    \put(0.42565551,0.85218122){\color[rgb]{0,0,0}\makebox(0,0)[lt]{\smash{\normalsize{$V$}}}}%
    \put(0.03052931,0.89069311){\color[rgb]{0,0,0}\makebox(0,0)[lt]{\smash{\normalsize{$\delta$}}}}%
  \end{picture}%
\caption{Kinematic single track model}%
\label{fig:singletrackmodel}%
\end{wrapfigure}
The vehicle is modeled physically with the nonlinear kinematic single track model \cite{Riekert1940}. See e.g. \cite{Schramm2014, Breuer2015, Rajamani2012} for further discussions of the model. Stationary steering is assumed. 
Key assumptions for this work are that the front and rear wheel pairs are modeled as a single wheel at each axle, see figure \ref{fig:singletrackmodel}. The mass of the vehicle $m_{veh}$ is concentrated at the center of gravity (CG). Moreover, roll dynamics and pitching are neglected. Aerodynamic forces are neither considered. While accelerating in a curve, the curve radius $R$, the steering angle $\delta$ and the side slip angle $\beta$ are time-invariant. This implies that the yaw rate $\dot{\alpha}$ varies. 

\vspace{1,5cm}
\subsubsection{Mathematical Model}
Based on the single track model, the equations of motion are constituted in \ref{sec:Derivation_of_equations}. From these, i.e. Eqs. \eqref{eq:Cons_Momentum} and \eqref{eq:Cons_Ang_Momentum}, expressions for $F_{\parallel}$, $F_{\perp}$ and $F_{\mu}$ are derived for the front and the rear axle, indicated with the subscripts $FA$ and $RA$, respectively. With Eq. \eqref{eq:Kamm_Circle2} the hazardous potential is stated for the front and the rear axle:
\begin{subequations}
\begin{equation}
P_{\mu,FA}= \frac{F_{\perp, FA}^{2}+F_{\parallel, FA}^{2}}{F_{\mu, FA}^{2}}, 
\label{eq:Worst_Case-eq-withlossoftraction_Front}
\end{equation}
\begin{equation}
P_{\mu,RA}= \frac{F_{\perp, RA}^{2}+F_{\parallel, RA}^{2}}{F_{\mu, RA}^{2}}, 
\label{eq:Worst_Case-eq-withlossoftraction_Rear}
\end{equation}
\end{subequations}
Expression for $F_{\parallel}$, $F_{\perp}$ and $F_{\mu}$ may be found in \ref{sec:Derivation_of_equations}.

Note that if a vehicle is a front-wheel drive or rear-wheel drive, $F_{\parallel, RA} = 0$, or  $F_{\parallel, FA} = 0$, respectively.

\subsubsection{Selecting the worst-case variant}\label{sec:Selecting_Vehicle_loss_traction}
Given \eqref{eq:Worst_Case-eq-withlossoftraction_Front}, \eqref{eq:Worst_Case-eq-withlossoftraction_Rear} a worst-case analysis of the vehicle parameters is performed.
Thereby, in case of undetermined variants, a worst-case parameter set is determined from which a worst case variant is constructed. 
Alternatively, given a set of variants with fixed parameters, the worst-case variant is selected. 
The worst-case parameter configuration or variant is the one for which  $P_{\mu,FA}$ or $P_{\mu,RA}$ is biggest. 
Note that a separate worst-case analysis is conducted for both cases: loss of traction of the front and rear axle, respectively.

\subsection{Use-Case ``{}Drivetrain fault entailing an unintended acceleration without loss of traction''{}}\label{sec:notractionloss_model}
\subsubsection{Use-case scenario assumptions}\label{sec:notractionloss_assumptions}

Maintaining traction implies that $P_{\mu,FA}< 1$ and $P_{\mu,RA}<1$, compare \eqref{eq:Worst_Case-eq-withlossoftraction_Front}, \eqref{eq:Worst_Case-eq-withlossoftraction_Rear}, respectively.
The worst-case driving scenario is driving straight ahead on a \textit{road with a normal friction coefficient}, i.e. $\mu = 1$. According to \cite{Sitkat-2015} this situation evaluates as $E4$. 
A drivetrain faults to allow the maximum engine torque. This fault entails an unintended acceleration. 
Note that, the mass of the vehicle is the curb weight.

\subsubsection{Characteristic of hazardous potential}
Given the scenario assumptions and description, the normalized accelerating potential of a vehicle 
\begin{equation}
P_{acc}=P_{acc, FA}+P_{acc, RA}
\label{eq:accelerationPotential}
\end{equation}
characterizes the hazardous potential of a variant. 

Similar to Section \ref{sec:tractionloss_model}, $P_{acc}$ directly correlates to the the acceleration, and the friction; it indirectly correlates to the kinetic energy: The bigger $P_{acc}$, the more kinetic energy is built up in a short period of time.
A vehicle with a higher acceleration overcomes distances to targets more swiftly. The time that remains for a driver, passenger or other road users to control the situation is shorter for a vehicle with a higher $P_{acc}$. Moreover, the kinetic energy in the moment of a mishap is generally bigger for bigger $P_{acc}$. 
Thus, $P_{acc}$ suits for characterizing the effects of the key vehicle parameters identified in e.g. \cite{Johansson2016, McGehee2000, ROSEN201125} to determine the hazardous potential, and correlating to the classification of S and C. 

\subsubsection{Physical Model}
We employ the physical model described in Section \ref{sec:physical_model}.
The steering angle $\delta$, the yaw rate $\dot{\alpha}$ and the side slip angle $\beta$ diminish.

\subsubsection{Mathematical Model}

The normalized accelerating potential of the front, and rear axle are 
\begin{subequations}
\begin{equation}
P_{acc, FA}=\frac{F_{\parallel, FA}}{F_{\mu, FA}} = \frac{T_{FA}~l}{r_{dyn}~ \mu ~g~m_{veh}~l_{RA}},
\label{eq:Potential_wheels_FA}
\end{equation}
\begin{equation}
P_{acc, RA}=\frac{F_{\parallel, RA}}{F_{\mu, RA}} = \frac{T_{RA}~l}{r_{dyn}~ \mu ~g~m_{veh}~l_{FA}},
\label{eq:Potential_wheels_RA}
\end{equation}
\end{subequations}
respectively. $F_{\parallel}$ and $F_{\mu}$ are derived for the front and the rear axle in \ref{sec:Derivation_of_equations}.
Hence, the accelerating potential of a variant with dedicated engines for each axle is 
\begin{equation}
P_{acc}=\frac{l}{r_{dyn}~ \mu ~g~m_{veh} }\frac{T_{mot, FA}~i_{FA}~l_{FA} + T_{mot, RA}~i_{RA}~l_{RA} }{\left( l_{FA}~ l_{RA}\right)}~.
\label{eq:accelerationPotentialvariant}
\end{equation}
In case a single engine is employed, the accelerating potential is 
\begin{equation}
P_{acc}=\frac{l~  T_{mot}}{r_{dyn}~ \mu ~g~m_{veh}}\frac{i_{tot,FA}~l_{FA} + i_{tot,RA}~l_{RA} }{\left( l_{FA}~ l_{RA}\right)}.
\label{eq:accelerationPotentialvariantsingleengine}
\end{equation}

\subsubsection{Selecting the worst-case variant}\label{sec:Vehicle_Selection_Acc_traction}
In case that variants are undetermined, the vehicle parameters are chosen to obtain the biggest $P_{acc}$. Thereby, the conditions $P_{\mu,FA}< 1$ and $P_{\mu,RA}<1$ need to be fulfilled. Note that $F_{\perp} = 0$.
From these parameters the worst-case variant is constructed.

In case the variants have been predetermined, for each variant the vehicle-parameters are set to maximize $P_{\mu,FA}$ and $P_{\mu,RA}$. Yet, the conditions $P_{\mu,FA}< 1$ and $P_{\mu,RA}<1$ need to be fulfilled. Note that $F_{\perp} = 0$. 
Proceeding, $P_{acc,FA}$, $P_{acc,RA}$ are evaluated according to Eqs. \eqref{eq:Potential_wheels_FA} and \eqref{eq:Potential_wheels_RA}, respectively, for each variant.
The worst-case variant is the one for which $P_{acc}$ is biggest.
\clearpage

\subsection{Application example}\label{Sec:application_example}

\begin{sidewaystable}
  \centering
  \begin{tabular}{|l|llllllll|}
\hline
\multicolumn{1}{|c|}{Variants:} &
  \multicolumn{1}{c}{\begin{tabular}[c]{@{}c@{}}B-Class \\ Sportcar 2017  \\ RWD\end{tabular}} &
  \multicolumn{1}{c}{\begin{tabular}[c]{@{}c@{}}D-Class \\ SUV v9 2017 \\ AWD\end{tabular}} &
  \multicolumn{1}{c}{\begin{tabular}[c]{@{}c@{}}F-Class \\ Sedan \\ AWD\end{tabular}} &
  \multicolumn{1}{c}{\begin{tabular}[c]{@{}c@{}}A-Class\\ Hatchback 2017\\ FWD\end{tabular}} &
	  \multicolumn{1}{c}{\begin{tabular}[c]{@{}c@{}}Large   \\European Van \\ RWD\end{tabular}} &
  \multicolumn{1}{c}{\begin{tabular}[c]{@{}c@{}}B-Class \\ Hatchback  2017\\ FWD\end{tabular}} &
  \multicolumn{1}{c}{\begin{tabular}[c]{@{}c@{}}European\\ Van \\ FWD\end{tabular}} &
  \multicolumn{1}{c|}{\begin{tabular}[c]{@{}c@{}}E-Class\\ Sedan 2017 \\ AWD\end{tabular}}\\ \hline
$T_{mot} [Nm]$   & 258    & 310   & 619   & 155   & 900   & 258   & 155   & 514    \\
$r_{dyn} [m]$    & 0.308  & 0.3575 & 0.3635 & 0.2915 & 0.4015 & 0.3105 & 0.3105 & 0.3815  \\
$i_{1st}$        & 3.358  & 3.358 & 4.596 & 3.78  & 3.1   & 3.358 & 3.78  & 4.38   \\
$i_{2nd}$        & 2.06   & 2.06  & 2.724 & 2.12  & 1.81  & 2.06  & 2.12  & 2.86   \\
$i_{3rd}$        & 1.404  & 1.404 & 1.864 & 1.36  & 1.41  & 1.404 & 1.36  & 1.92   \\
$i_{4th}$        & 1      & 1     & 1.464 & 1.03  & 1     & 1     & 1.03  & 1.37   \\
$i_{5th}$        & 0.713  & 0.713 & 1.231 & 0.84  & 0.71  & 0.713 & 0.84  & 1      \\
$i_{6th}$        & 0.582  & 0.582 & 1     & 0     & 0.61  & 0.528 & 0     & 0.82   \\
$i_{7th}$        & 0      & 0     & 0.824 & 0     & 0     & 0     & 0     & 0.7   \\
$i_{Diff,FA}$    & 0      & 1.64   & 1.06  & 4.1   & 0     & 4.1   & 4.1   & 1.06   \\
$i_{Diff,RA}$    & 4.1    & 2.46   & 1.59  & 0     & 4.1   & 0     & 0     & 1.59   \\
$l_{FA} [m]$     & 1.186  & 1.05  & 1.265 & 1.1   & 1.35  & 1.04  & 1.35  & 1.4    \\
$l_{RA} [m]$     & 1.144  & 1.61  & 1.895 & 1.25  & 1.75  & 1.56  & 1.23  & 1.65   \\
$m_{veh} [kg]$   & 1140   & 1610  & 2220  & 833   & 2400  & 1230  & 1300  & 1830   \\
\hline
\end{tabular}
  \caption{Example variants, adopted from \cite{CarSimNotes2018}}
  \label{tab:Vehicles_part1}
\end{sidewaystable}

\begin{sidewaystable}
  \centering
  \begin{tabular}{|l|lllll|}
\hline
\multicolumn{1}{|c|}{Variants:} &
   \multicolumn{1}{c}{\begin{tabular}[c]{@{}c@{}}E-Class\\ SUV 2017      \\ AWD\end{tabular}} &
  \multicolumn{1}{c}{\begin{tabular}[c]{@{}c@{}}C-Class\\ Hatchback 2017\\ FWD\end{tabular}} &
  \multicolumn{1}{c}{\begin{tabular}[c]{@{}c@{}}D-Class\\ Sedan 2017    \\ FWD\end{tabular}} &
  \multicolumn{1}{c}{\begin{tabular}[c]{@{}c@{}}D-Class\\ Minivan 2017  \\ FWD\end{tabular}} &
  \multicolumn{1}{c|}{\begin{tabular}[c]{@{}c@{}}SUV\\ full size \\ AWD\end{tabular}} 
	\\ \hline
$T_{mot} [Nm]$  & 413    & 258   & 310   & 310   & 514   \\
$r_{dyn} [m]$    & 0.402 & 0.334 & 0.334 & 0.3685 & 0.407 \\
$i_{1st}$        & 3.358 & 3.358 & 3.358 & 3.358 & 4.38  \\
$i_{2nd}$        & 2.06  & 2.06  & 2.06  & 2.06  & 2.86  \\
$i_{3rd}$        & 1.404 & 1.404 & 1.404 & 1.404 & 1.92  \\
$i_{4th}$        & 1     & 1     & 1     & 1     & 1.37  \\
$i_{5th}$        & 0.713 & 0.713 & 0.713 & 0.713 & 1     \\
$i_{6th}$        & 0.582 & 0.582 & 0.582 & 0.582 & 0.82  \\
$i_{7th}$        & 0     & 0     & 0     & 0     & 0.7   \\
$i_{Diff,FA}$    & 1.64   & 4.1   & 4.1   & 4.1   & 1.06  \\
$i_{Diff,RA}$    & 2.46   & 0     & 0     & 0     & 1.59  \\
$l_{FA} [m]$     & 1.18  & 1.015 & 1.11  & 1.35  & 1.33  \\
$l_{RA} [m]$     & 1.77  & 1.895 & 1.67  & 1.65  & 1.81  \\
$m_{veh} [kg]$   & 1860  & 1412  & 1530  & 2000  & 2532  \\
\hline
\end{tabular}
  \caption{Example variants, adopted from \cite{CarSimNotes2018}, continued}
  \label{tab:Vehicles_part2}
\end{sidewaystable}

As an application example, the variants listed in tables \ref{tab:Vehicles_part1} and \ref{tab:Vehicles_part2} are studied for their potential to constitute representative variants for use-cases of a
drivetrain fault entailing an unintended acceleration with loss of traction at front, rear axle, respectively, and drivetrain fault entailing an unintended acceleration with loss of traction at rear axle. Note that for the AWD variants, the torque distribution to the front and rear axle are fixed and at a ratio of $3$ to $2$, resulting in $i_{Diff,FA}$, $i_{Diff,RA}$. 

\subsubsection{Use-Case ``{}Drivetrain fault entailing an unintended acceleration with loss of traction at front axle''{}}

\begin{table}%
\centering
  \begin{tabular}{|l|ccccc|}
\hline
\multicolumn{1}{|c|}{Variants:} &
  \multicolumn{1}{c}{\begin{tabular}[c]{@{}c@{}}B-Class \\ Sportcar 2017  \\ RWD\end{tabular}} &
  \multicolumn{1}{c}{\begin{tabular}[c]{@{}c@{}}D-Class \\ SUV v9 2017 \\ AWD\end{tabular}} &
  \multicolumn{1}{c}{\begin{tabular}[c]{@{}c@{}}F-Class \\ Sedan  \\ AWD\end{tabular}} &
  \multicolumn{1}{c}{\begin{tabular}[c]{@{}c@{}}A-Class\\ Hatchback 2017\\ FWD\end{tabular}} &
	\multicolumn{1}{c|}{\begin{tabular}[c]{@{}c@{}}Large   \\European Van \\ RWD\end{tabular}} 
	\\ \hline
	$P_{\mu,FA}$  & 0.714    & 1.560   & 2.729  & 26.801 & 0.909  \\ 
	\hline \hline
	\multicolumn{1}{|c|}{Variants:} &
	\multicolumn{1}{c}{\begin{tabular}[c]{@{}c@{}}B-Class \\ Hatchback  2017\\ FWD\end{tabular}} &
  \multicolumn{1}{c}{\begin{tabular}[c]{@{}c@{}}European\\ Van \\ FWD\end{tabular}} &
  \multicolumn{1}{c}{\begin{tabular}[c]{@{}c@{}}E-Class\\ Sedan 2017 \\ AWD\end{tabular}} &
	\multicolumn{1}{c}{\begin{tabular}[c]{@{}c@{}}E-Class\\ SUV 2017      \\ AWD\end{tabular}} &
  \multicolumn{1}{c|}{\begin{tabular}[c]{@{}c@{}}C-Class\\ Hatchback 2017\\ FWD\end{tabular}}
	\\ \hline
$P_{\mu,FA}$  & 19.233    & 11.768   & 2.758  & 1.685  & 10.933 \\
\hline \hline
\multicolumn{1}{|c|}{Variants:} &
\multicolumn{1}{c}{\begin{tabular}[c]{@{}c@{}}D-Class\\ Sedan 2017    \\ FWD\end{tabular}} &
  \multicolumn{1}{c}{\begin{tabular}[c]{@{}c@{}}D-Class\\ Minivan 2017  \\ FWD\end{tabular}} &
  \multicolumn{1}{c}{\begin{tabular}[c]{@{}c@{}}SUV\\ full size \\ AWD\end{tabular}} & & \\
	\hline
	$P_{\mu,FA}$  & 15.797    & 8.823   & 1.052   & &\\
\hline
\end{tabular}
  \caption{$P_{\mu,FA}$ for example variants, use-case ``{}Drivetrain fault entailing an unintended acceleration with loss of traction at front axle''{}.}
  \label{tab:Vehicles_loss_FA}
\end{table}

Given the set of variants and their parameters, and the use-case scenario as detailed in Section \ref{sec:use-case-assumptions-lossoftraction}, Eq. \eqref{eq:Worst_Case-eq-withlossoftraction_Front} is evaluated for each variant. The procedure described in Section \ref{sec:Selecting_Vehicle_loss_traction} holds. The results are tabulated in Table \ref{tab:Vehicles_loss_FA}.
Thereby, for all variants, the first gear is selected to maximize $i_{tot}$. 

One finds that the vehicle \textit{A-Class Hatchback 2017, FWD}, see table \ref{tab:Vehicles_part1} represents the worst-case variant for the case of a drivetrain fault entailing an unintended acceleration with loss of traction of the front axle.
\clearpage
\subsubsection{Use-Case ``{}Drivetrain fault entailing an unintended acceleration with loss of traction at rear axle''{}}
\begin{table}%
\centering
  \begin{tabular}{|l|ccccc|}
\hline
\multicolumn{1}{|c|}{Variants:} &
  \multicolumn{1}{c}{\begin{tabular}[c]{@{}c@{}}B-Class \\ Sportcar 2017  \\ RWD\end{tabular}} &
  \multicolumn{1}{c}{\begin{tabular}[c]{@{}c@{}}D-Class \\ SUV v9 2017 \\ AWD\end{tabular}} &
  \multicolumn{1}{c}{\begin{tabular}[c]{@{}c@{}}F-Class \\ Sedan  \\ AWD\end{tabular}} &
  \multicolumn{1}{c}{\begin{tabular}[c]{@{}c@{}}A-Class\\ Hatchback 2017\\ FWD\end{tabular}} &
	\multicolumn{1}{c|}{\begin{tabular}[c]{@{}c@{}}Large   \\European Van \\ RWD\end{tabular}} 
	\\ \hline
	$P_{\mu,RA}$  & 26.175    & 8.772   & 13.255  & 0.521 & 48.781  \\ 
	\hline \hline
	\multicolumn{1}{|c|}{Variants:} &
	\multicolumn{1}{c}{\begin{tabular}[c]{@{}c@{}}B-Class \\ Hatchback  2017\\ FWD\end{tabular}} &
  \multicolumn{1}{c}{\begin{tabular}[c]{@{}c@{}}European\\ Van \\ FWD\end{tabular}} &
  \multicolumn{1}{c}{\begin{tabular}[c]{@{}c@{}}E-Class\\ Sedan 2017 \\ AWD\end{tabular}} &
	\multicolumn{1}{c}{\begin{tabular}[c]{@{}c@{}}E-Class\\ SUV 2017      \\ AWD\end{tabular}} &
  \multicolumn{1}{c|}{\begin{tabular}[c]{@{}c@{}}C-Class\\ Hatchback 2017\\ FWD\end{tabular}}
	\\ \hline
$P_{\mu,RA}$  & 0.521   & 0.521   & 8.625  & 8.972  & 0.521 \\
\hline \hline
\multicolumn{1}{|c|}{Variants:} &
\multicolumn{1}{c}{\begin{tabular}[c]{@{}c@{}}D-Class\\ Sedan 2017    \\ FWD\end{tabular}} &
  \multicolumn{1}{c}{\begin{tabular}[c]{@{}c@{}}D-Class\\ Minivan 2017  \\ FWD\end{tabular}} &
  \multicolumn{1}{c}{\begin{tabular}[c]{@{}c@{}}SUV\\ full size \\ AWD\end{tabular}} & & \\
	\hline
	$P_{\mu,RA}$  & 0.521    & 0.521   & 4.889   & &\\
\hline
\end{tabular}
  \caption{$P_{\mu,RA}$ for example variants, use-case ``{}Drivetrain fault entailing an unintended acceleration with loss of traction at rear axle''{}.}
  \label{tab:Vehicles_loss_RA}
\end{table}

Given the set of variants and their parameters, and the use-case scenario as detailed in Section \ref{sec:use-case-assumptions-lossoftraction}, Eq. \eqref{eq:Worst_Case-eq-withlossoftraction_Rear} is evaluated for each variant. The procedure described in Section \ref{sec:Selecting_Vehicle_loss_traction} holds. The results are listed in Table \ref{tab:Vehicles_loss_FA}.
Thereby, for all variants, the first gear is selected to maximize $i_{tot}$.

One finds that the vehicle \textit{Large European Van, RWD}, see table \ref{tab:Vehicles_part1}, represents the worst-case variant for the case of a drivetrain fault entailing an unintended acceleration with loss of traction of the rear axle.
\clearpage
\subsubsection{Use-Case ``{}Drivetrain fault entailing an unintended acceleration without loss of traction''{}}

\begin{table}%
\centering
  \begin{tabular}{|l|ccccc|}
\hline
\multicolumn{1}{|c|}{Variants:} &
  \multicolumn{1}{c}{\begin{tabular}[c]{@{}c@{}}B-Class \\ Sportcar 2017  \\ RWD\end{tabular}} &
  \multicolumn{1}{c}{\begin{tabular}[c]{@{}c@{}}D-Class \\ SUV v9 2017 \\ AWD\end{tabular}} &
  \multicolumn{1}{c}{\begin{tabular}[c]{@{}c@{}}F-Class \\ Sedan  \\ AWD\end{tabular}} &
  \multicolumn{1}{c}{\begin{tabular}[c]{@{}c@{}}A-Class\\ Hatchback 2017\\ FWD\end{tabular}} &
	\multicolumn{1}{c|}{\begin{tabular}[c]{@{}c@{}}Large   \\European Van \\ RWD\end{tabular}} 
	\\ \hline
	gear    & $3rd$    			& $2nd$   & $2nd$  	& $3rd$ & $4th$  \\
	$P_{\mu,FA}$  	& 0    	& 0.094   & 0.142		& 0.465 & 0  \\
	$P_{\mu,RA}$  	& 0.718 & 0.497   & 0.716  	& 0 		& 0.803  \\ 
	$P_{acc}$  			& 0.847 & 1.011   & 1.222  	& 0.682 & 0.896  \\ 
	\hline \hline
	\multicolumn{1}{|c|}{Variants:} &
	\multicolumn{1}{c}{\begin{tabular}[c]{@{}c@{}}B-Class \\ Hatchback  2017\\ FWD\end{tabular}} &
  \multicolumn{1}{c}{\begin{tabular}[c]{@{}c@{}}European\\ Van \\ FWD\end{tabular}} &
  \multicolumn{1}{c}{\begin{tabular}[c]{@{}c@{}}E-Class\\ Sedan 2017 \\ AWD\end{tabular}} &
	\multicolumn{1}{c}{\begin{tabular}[c]{@{}c@{}}E-Class\\ SUV 2017      \\ AWD\end{tabular}} &
  \multicolumn{1}{c|}{\begin{tabular}[c]{@{}c@{}}C-Class\\ Hatchback 2017\\ FWD\end{tabular}}
	\\ \hline
	gear    			& $2nd$ & $2nd$ & $2nd$  & $2nd$ & $2nd$  \\
	$P_{\mu,FA}$  & 0.940 & 0.509 & 0.177  & 0.101 & 0.523  \\
	$P_{\mu,RA}$  & 0    	& 0   	& 0.553  & 0.509 & 0  \\ 
	$P_{acc}$  		& 0.969 & 0.714 & 1.164  & 1.030 & 0.723  \\ 
\hline \hline
\multicolumn{1}{|c|}{Variants:} &
\multicolumn{1}{c}{\begin{tabular}[c]{@{}c@{}}D-Class\\ Sedan 2017    \\ FWD\end{tabular}} &
  \multicolumn{1}{c}{\begin{tabular}[c]{@{}c@{}}D-Class\\ Minivan 2017  \\ FWD\end{tabular}} &
  \multicolumn{1}{c}{\begin{tabular}[c]{@{}c@{}}SUV\\ full size \\ AWD\end{tabular}} & & \\
	\hline
	gear    			& $2nd$ & $2nd$ & $1st$ & &   \\
	$P_{\mu,FA}$  & 0.756 & 0.434 & 0.168 & &  \\
	$P_{\mu,RA}$  & 0    	& 0  		& 0.699 & & \\ 
	$P_{acc}$  		& 0.869 & 0.658 & 1.245 & &  \\ 
\hline
\end{tabular}
  \caption{Worst-case analysis for example variants, use-case ``{}Drivetrain fault entailing an unintended acceleration without loss of traction''{}.}
  \label{tab:Vehicles_traction}
\end{table}

Given the set of variants and their parameters, and the use-case scenario as detailed in Section \ref{sec:notractionloss_assumptions}, the worst-case $P_{acc}$, see Eq. \eqref{eq:accelerationPotential}, is evaluated for each variant. Thereby, the procedure of Section \ref{sec:Vehicle_Selection_Acc_traction} holds. 
The results are listed in Tables \ref{tab:Vehicles_traction}. Also, the gear, the corresponding total gear ratios for the front and rear axle, $i_{tot,FA}$ and $i_{tot,RA}$, $P_{\mu,FA}$ and $P_{\mu,RA}$, $P_{acc,FA}$, $P_{acc,RA}$ are listed. 

One finds that the vehicle \textit{SUV full size, AWD}, see table \ref{tab:Vehicles_part2}, represents the variant for the case of a drivetrain fault entailing an unintended acceleration without loss of traction.

\clearpage
\section{Conclusion}\label{sec:Conclusion}
In this work we have found that a well-defined item definition is the key to limit the effort for functional safety development. We have pointed out that in platform development common requirements allow clustering variants into a single item definition. 
A method to define cluster-representative variants for functional safety development according to \cite{iso26262_2nd} has been developed. As a prerequisite the variants listed in the item definition are to be put in a context, i.e. a use-case situation. To comply with ISO 26262:2018 \cite{iso26262_2nd} both, a representative variant, defined by its configuration parameters, and the scenario conditions need to adhere to the worst-case assumption. 
A key step in efficiently deriving a use-case specific representative worst-case variant is to state a characteristic that relates parameters of the item to the hazard-source potential of the item. 
Assuming the use-case scenario, a physical model shall be stated such to focus the item parameters to the most relevant ones. Subsequent mathematical modeling specifies the characteristic in terms of item parameters. An extreme value analysis leads to the sought worst-case, representative variant. It is to be used for all the subsequent steps of functional safety development.

The method has been applied to assess two use-case situations in drivetrain functional safety development: ``{}Drivetrain fault entailing an unintended acceleration with loss of traction''{} and ``{}Drivetrain fault entailing an unintended acceleration without loss of traction''{}. For the two use-case situations characteristics have been proposed that may be used for functional safety development.  

\section*{Acknowledgments}
The authors acknowledge their colleagues of techcos and the Technische University of Munich for invigorating discussions. Elisa Schranner is acknowledged for constructive criticism of the manuscript. The Dean of Mechanical Engineering at the Technische University of Munich, Nikolaus Adams, is acknowledged for supporting the first author to pursue this work.  
\clearpage
\appendix

\section{Derivation of mathematical model for considered drivetrain use-cases}\label{sec:Derivation_of_equations}
\subsection{Configuration}
An accelerating vehicle, modeled with the nonlinear kinematic single track model \cite{Riekert1940}, driving a curve of radius $R$ at a momentary velocity $V$, see Section \ref{sec:DriveTrain-WC}, is considered.
\begin{figure}[ht]%
\centering
\subfloat[x-y plane]{
\def\svgwidth{0.45\columnwidth}
  \makeatletter%
  \providecommand\color[2][]{%
    \errmessage{(Inkscape) Color is used for the text in Inkscape, but the package 'color.sty' is not loaded}%
    \renewcommand\color[2][]{}%
  }%
  \providecommand\transparent[1]{%
    \errmessage{(Inkscape) Transparency is used (non-zero) for the text in Inkscape, but the package 'transparent.sty' is not loaded}%
    \renewcommand\transparent[1]{}%
  }%
  \providecommand\rotatebox[2]{#2}%
	\setlength{\unitlength}{\svgwidth}%
  \global\let\svgwidth\undefined%
  \global\let\svgscale\undefined%
  \makeatother%
  \begin{picture}(1,0.95683449)%
    \setlength\tabcolsep{0pt}%
    \put(0,0){\includegraphics[width=\unitlength,page=1]{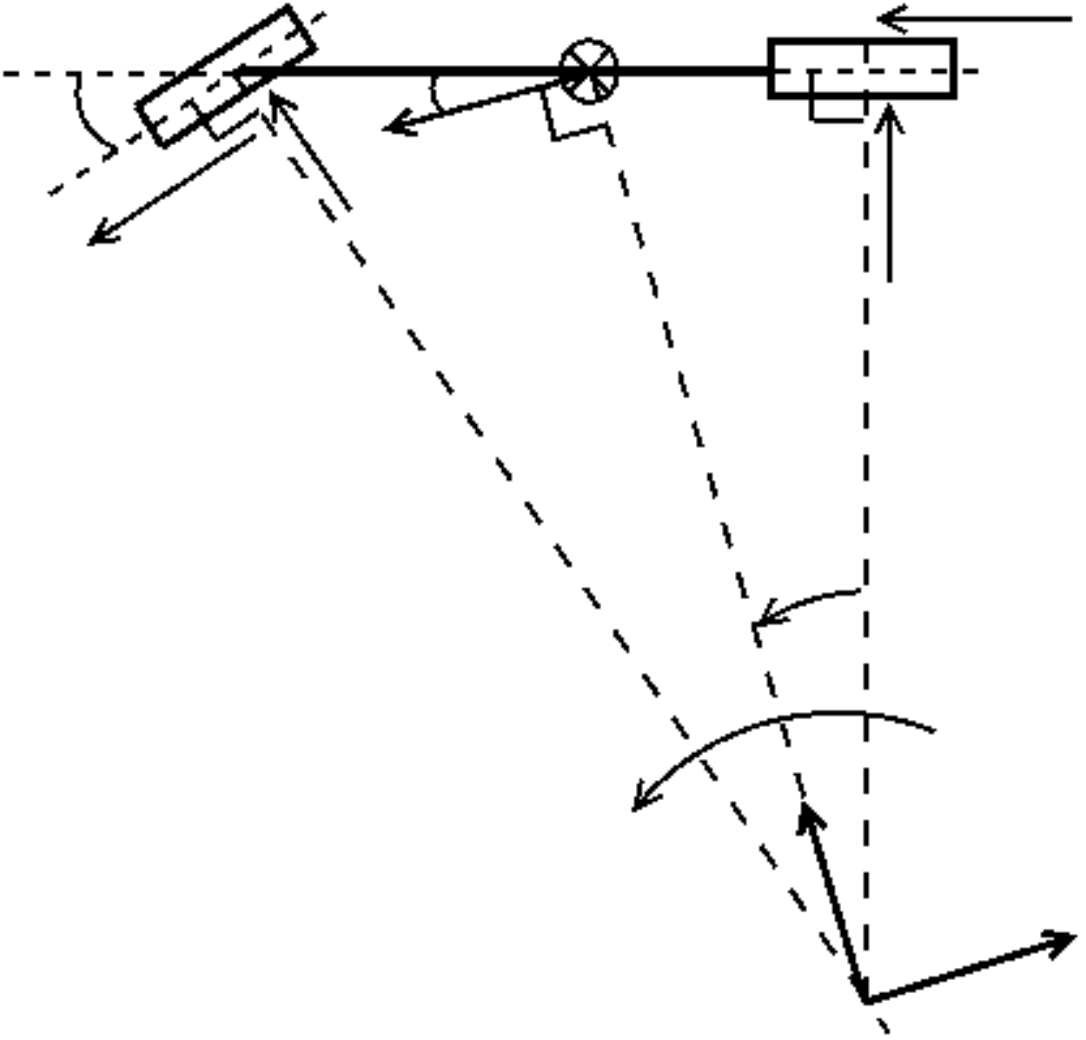}}%
    \put(0.92779001,0.11869891){\color[rgb]{0,0,0}\makebox(0,0)[lt]{\smash{\normalsize{$x_I$}}}}%
    \put(0.75209341,0.2328427){\color[rgb]{0,0,0}\makebox(0,0)[lt]{\smash{\normalsize{$y_I$}}}}%
    \put(0.38514721,0.90721834){\color[rgb]{0,0,0}\makebox(0,0)[lt]{\smash{\normalsize{$\beta$}}}}%
    \put(0.63235656,0.30314491){\color[rgb]{0,0,0}\makebox(0,0)[lt]{\smash{\normalsize{$\dot{\alpha}$}}}}%
    \put(0.81456093,0.48588623){\color[rgb]{0,0,0}\makebox(0,0)[lt]{\smash{\normalsize{$R'$}}}}%
    \put(0.71299205,0.43257899){\color[rgb]{0,0,0}\makebox(0,0)[lt]{\smash{\normalsize{$\beta$}}}}%
    \put(0.65723669,0.55128392){\color[rgb]{0,0,0}\makebox(0,0)[lt]{\smash{\normalsize{$R$}}}}%
    \put(0.53313594,0.93977314){\color[rgb]{0,0,0}\makebox(0,0)[lt]{\smash{\normalsize{$CG$}}}}%
    \put(0.18706277,0.96495297){\color[rgb]{0,0,0}\makebox(0,0)[lt]{\smash{\normalsize{$FA$}}}}%
    \put(0.71229266,0.96675156){\color[rgb]{0,0,0}\makebox(0,0)[lt]{\smash{\normalsize{$RA$}}}}%
    \put(0.40773634,0.80428153){\color[rgb]{0,0,0}\makebox(0,0)[lt]{\smash{\normalsize{$V$}}}}%
    \put(0.0340353,0.82286681){\color[rgb]{0,0,0}\makebox(0,0)[lt]{\smash{\normalsize{$\delta$}}}}%
    \put(0.87316395,0.96735109){\color[rgb]{0,0,0}\makebox(0,0)[lt]{\smash{\normalsize{$F_{\parallel, RA}$}}}}%
    \put(0.13580386,0.70422405){\color[rgb]{0,0,0}\makebox(0,0)[lt]{\smash{\normalsize{$F_{\parallel, FA}$}}}}%
    \put(0.8328963,0.77616657){\color[rgb]{0,0,0}\makebox(0,0)[lt]{\smash{\normalsize{$F_{\perp, RA}$}}}}%
    \put(0.33154726,0.73839676){\color[rgb]{0,0,0}\makebox(0,0)[lt]{\smash{\normalsize{$F_{\perp, FA}$}}}}%
  \end{picture}%
\label{fig:Singletrack_top-view}}
\subfloat[x-z plane]{
\def\svgwidth{0.45\columnwidth}
  \makeatletter%
  \providecommand\color[2][]{%
    \errmessage{(Inkscape) Color is used for the text in Inkscape, but the package 'color.sty' is not loaded}%
    \renewcommand\color[2][]{}%
  }%
  \providecommand\transparent[1]{%
    \errmessage{(Inkscape) Transparency is used (non-zero) for the text in Inkscape, but the package 'transparent.sty' is not loaded}%
    \renewcommand\transparent[1]{}%
  }%
  \providecommand\rotatebox[2]{#2}%
  \setlength{\unitlength}{\svgwidth}%
  \global\let\svgwidth\undefined%
  \global\let\svgscale\undefined%
  \makeatother%
  \begin{picture}(1,0.44807123)%
    \setlength\tabcolsep{0pt}%
    \put(0,0){\includegraphics[width=\unitlength,page=1]{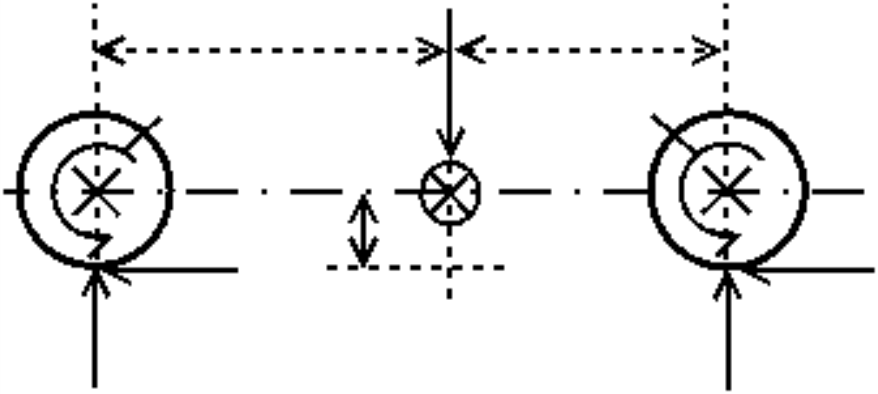}}%
    \put(0.55188252,0.16709154){\color[rgb]{0,0,0}\makebox(0,0)[lt]{\smash{\normalsize{$CG$}}}}%
    \put(0.29757859,0.40936491){\color[rgb]{0,0,0}\makebox(0,0)[lt]{\smash{\normalsize{$l_{FA}$}}}}%
    \put(0.61447286,0.40881627){\color[rgb]{0,0,0}\makebox(0,0)[lt]{\smash{\normalsize{$l_{RA}$}}}}%
    \put(0.89427645,0.30926035){\color[rgb]{0,0,0}\makebox(0,0)[lt]{\smash{\normalsize{$RA$}}}}%
    \put(0.01681649,0.32908994){\color[rgb]{0,0,0}\makebox(0,0)[lt]{\smash{\normalsize{$FA$}}}}%
    \put(0.54411396,0.28643135){\color[rgb]{0,0,0}\makebox(0,0)[lt]{\smash{\normalsize{$F_{N}$}}}}%
    \put(0.12431492,0.01995345){\color[rgb]{0,0,0}\makebox(0,0)[lt]{\smash{\normalsize{$F_{N,FA}$}}}}%
    \put(0.84709916,0.02029742){\color[rgb]{0,0,0}\makebox(0,0)[lt]{\smash{\normalsize{$F_{N,RA}$}}}}%
    \put(0.30922548,0.18200558){\color[rgb]{0,0,0}\makebox(0,0)[lt]{\smash{\normalsize{$r_{dyn}$}}}}%
    \put(0.64797125,0.31792418){\color[rgb]{0,0,0}\makebox(0,0)[lt]{\smash{\normalsize{$T_{RA}$}}}}%
    \put(0.18310872,0.30102691){\color[rgb]{0,0,0}\makebox(0,0)[lt]{\smash{\normalsize{$T_{FA}$}}}}%
    \put(0.16566786,0.08965032){\color[rgb]{0,0,0}\makebox(0,0)[lt]{\smash{\normalsize{$F_{\parallel,FA}$}}}}%
    \put(0.8958707,0.09090402){\color[rgb]{0,0,0}\makebox(0,0)[lt]{\smash{\normalsize{$F_{\parallel,RA}$}}}}%
  \end{picture}%
\label{fig:Singletrack_side-view}}
\caption{kinematic single track model with forces acting on vehicle. } 
\label{fig:Vehicle_modeling_Details}
\end{figure}
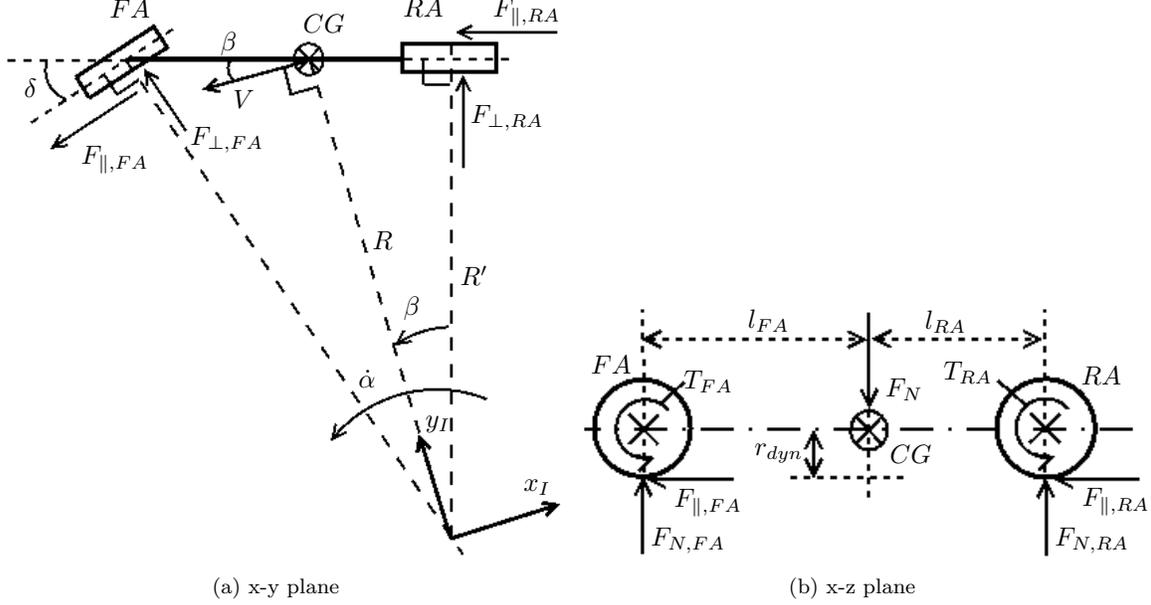
The vehicle is depicted in the x-y-plane and x-z-plane in Fig. \ref{fig:Singletrack_top-view} and Fig. \ref{fig:Singletrack_side-view}, respectively. The forces acting on the wheels/axle are shown. 
$F_{\parallel}$ and $F_{\perp}$ are the longitudinal (accelerating) and transversal forces acting on the wheels. $T_{FA}$, $T_{RA}$ denote the front- and rear-wheel/axle torque, $r_{dyn}$ is the dynamic wheel radius, which, due to assuming stationarity and neglecting pitching, is identical for the front and rear wheels and time invariant. $F_N$, $F_{N, FA}$ and $F_{N, RA}$ are the normal forces. The normal force $F_N= m_{veh} g$, with $g=9,81 \frac{m}{s^2}$, $m_{veh}$ denoting the gravitational constant and the vehicle mass. Also, $l = l_{FA}+l_{RA}$ is the wheel base.
The Ackermann turning angle and side slip angle are approximated as
\begin{equation}
tan \delta = \frac{l}{R'} = \frac{l}{\sqrt{R^2 - l_{RA}^2}}~, 
\label{eq:Ackermann_angle}
\end{equation}
\begin{equation}
tan \beta = \frac{l_{RA}}{R'} = \frac{l_{RA}}{\sqrt{R^2 - l_{RA}^2}} 
\label{eq:sideslip_angle}
\end{equation}

\begin{figure}[ht]
\centering
\subfloat[Inertial (subscript I) and vehicle-body (subscript B) coordinate systems]{
\def\svgwidth{0.45\columnwidth}
  \makeatletter%
  \providecommand\color[2][]{%
    \errmessage{(Inkscape) Color is used for the text in Inkscape, but the package 'color.sty' is not loaded}%
    \renewcommand\color[2][]{}%
  }%
  \providecommand\transparent[1]{%
    \errmessage{(Inkscape) Transparency is used (non-zero) for the text in Inkscape, but the package 'transparent.sty' is not loaded}%
    \renewcommand\transparent[1]{}%
  }%
  \providecommand\rotatebox[2]{#2}%
	\setlength{\unitlength}{\svgwidth}%
  \global\let\svgwidth\undefined%
  \global\let\svgscale\undefined%
  \makeatother%
  \begin{picture}(1,0.64675317)%
    \setlength\tabcolsep{0pt}%
    \put(0,0){\includegraphics[width=\unitlength,page=1]{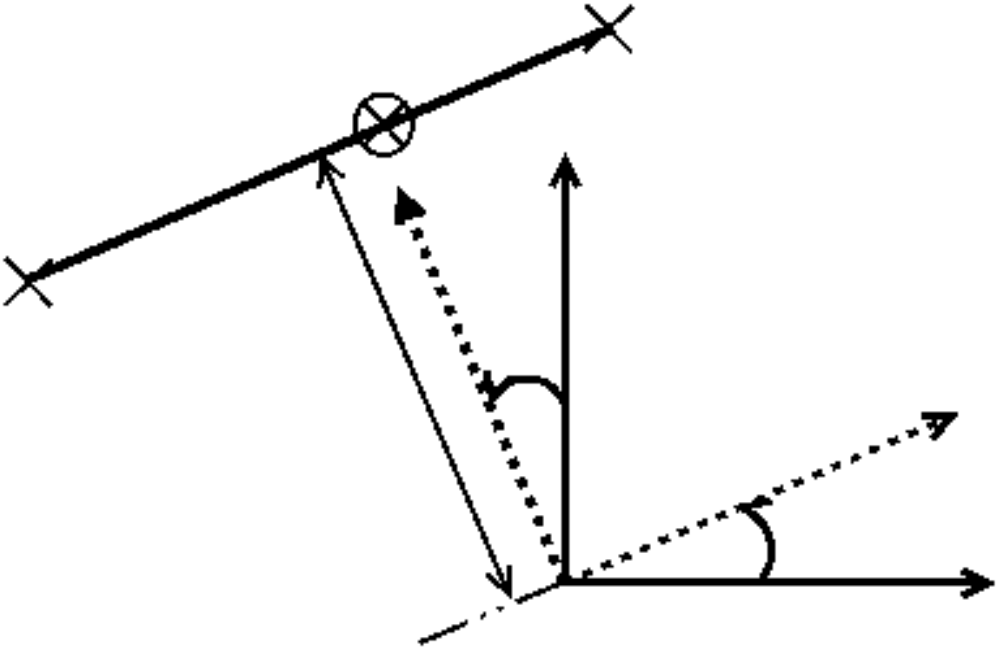}}%
    \put(0.59534815,0.48615731){\color[rgb]{0,0,0}\makebox(0,0)[lt]{\smash{\normalsize{$y_I$}}}}%
    \put(0.97480144,0.09275992){\color[rgb]{0,0,0}\makebox(0,0)[lt]{\smash{\normalsize{$x_I$}}}}%
    \put(0.92557781,0.25394432){\color[rgb]{0,0,0}\makebox(0,0)[lt]{\smash{\normalsize{$x_B$}}}}%
    \put(0.43939975,0.44244148){\color[rgb]{0,0,0}\makebox(0,0)[lt]{\smash{\normalsize{$y_B$}}}}%
    \put(0.78448329,0.09550281){\color[rgb]{0,0,0}\makebox(0,0)[lt]{\smash{\normalsize{$\beta$}}}}%
    \put(0.49770222,0.29497147){\color[rgb]{0,0,0}\makebox(0,0)[lt]{\smash{\normalsize{$\beta$}}}}%
    \put(0.34371329,0.23448897){\color[rgb]{0,0,0}\makebox(0,0)[lt]{\smash{\normalsize{$R$}}}}%
    \put(0.55772074,0.65601033){\color[rgb]{0,0,0}\makebox(0,0)[lt]{\smash{\normalsize{$RA$}}}}%
    \put(-0.02186327,0.40814386){\color[rgb]{0,0,0}\makebox(0,0)[lt]{\smash{\normalsize{$FA$}}}}%
    \put(0.34239921,0.57400658){\color[rgb]{0,0,0}\makebox(0,0)[lt]{\smash{\normalsize{$CG$}}}}%
  \end{picture}%
\label{fig:I-B-System}
}\hspace{0.1\columnwidth}
\subfloat[Vehicle-body (subscript B) and front axle (subscript FA) coordinate systems]{
\def\svgwidth{0.3\columnwidth}
  \makeatletter%
  \providecommand\color[2][]{%
    \errmessage{(Inkscape) Color is used for the text in Inkscape, but the package 'color.sty' is not loaded}%
    \renewcommand\color[2][]{}%
  }%
  \providecommand\transparent[1]{%
    \errmessage{(Inkscape) Transparency is used (non-zero) for the text in Inkscape, but the package 'transparent.sty' is not loaded}%
    \renewcommand\transparent[1]{}%
  }%
  \providecommand\rotatebox[2]{#2}%
  \setlength{\unitlength}{\svgwidth}%
  \global\let\svgwidth\undefined%
  \global\let\svgscale\undefined%
  \makeatother%
  \begin{picture}(1,0.75744682)%
    \setlength\tabcolsep{0pt}%
    \put(0,0){\includegraphics[width=\unitlength,page=1]{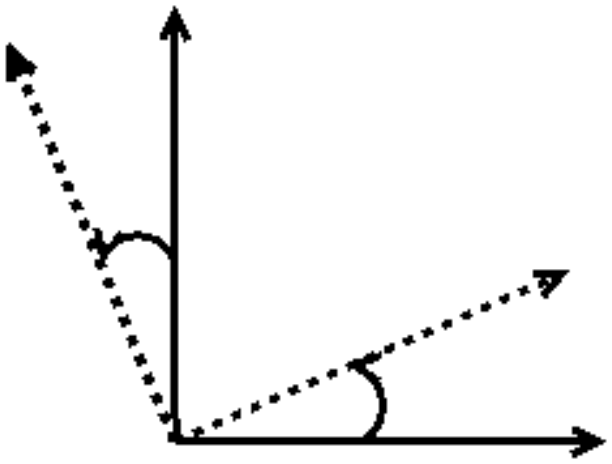}}%
    \put(0.31945312,0.76170233){\color[rgb]{0,0,0}\makebox(0,0)[lt]{\smash{\normalsize{$y_B$}}}}%
    \put(0.91500415,0.07834657){\color[rgb]{0,0,0}\makebox(0,0)[lt]{\smash{\normalsize{$x_B$}}}}%
    \put(0.8912962,0.33974474){\color[rgb]{0,0,0}\makebox(0,0)[lt]{\smash{\normalsize{$x_{FA}$}}}}%
    \put(-0.00359954,0.70691818){\color[rgb]{0,0,0}\makebox(0,0)[lt]{\smash{\normalsize{$y_{FA}$}}}}%
    \put(0.66758825,0.09536823){\color[rgb]{0,0,0}\makebox(0,0)[lt]{\smash{\normalsize{$\delta$}}}}%
    \put(0.19909776,0.39297754){\color[rgb]{0,0,0}\makebox(0,0)[lt]{\smash{\normalsize{$\delta$}}}}%
  \end{picture}%
\label{fig:B-FA-System}}
\caption{Coordinate systems } 
\label{fig:Coordinate_systems}
\end{figure}
The inertial (I), vehicle-body (B) and front axle (FA) coordinate systems are depicted in Fig. \ref{fig:Coordinate_systems}. $R(\dot{R},\ddot{R})$ and $\beta(\dot{\beta},\ddot{\beta})$, are the curve radius and angular rotation of the vehicle body coordinate system with respect to the inertial system, respectively, see Fig. \ref{fig:I-B-System}. The FA-system is rotated about the B-system by $\delta$, see Fig. \ref{fig:B-FA-System}.

\subsection{Transformation of coordinate systems}
Following rotational coordinate transformations are relevant:

\begin{multicols}{2}

$\underline{r}_{FA}=\underline{\underline{A}}_{F,B}\underline{r}_{B} =
\left( {\begin{array}{*{20}c}
    cos \delta & sin \delta & 0 \\
   -sin \delta & cos \delta  & 0\\
	0 & 0 & 1 \\
 \end{array} } \right) \underline{r}_{B}$
 
$\underline{r}_{B}=\underline{\underline{A}}_{B,I}\underline{r}_{I}  \quad =
\left( {\begin{array}{*{20}c}
    cos \beta & -sin \beta & 0  \\
   sin \beta &  cos \beta & 0  \\
	0 & 0 & 1 \\
 \end{array} } \right) \underline{r}_{I}$

 \columnbreak
$\underline{r}_{I} \textrm{: position vector in I-System}$

$\underline{r}_{B} \textrm{: position vector in B-System}$

$\underline{r}_{FA} \textrm{: position vector in FA-System}$

\end{multicols}

\subsection{Forces acting on the wheels / axles}
\noindent\underline{Position vector CG (center of gravity), FA, RA in I-, B-System}\\
The coordinate vectors for the CG, the front- and rear-axle are, see Fig. \ref{fig:Singletrack_top-view}: 
\begin{equation*}
\begin{split}
 \underline{x}_{I,\overrightarrow{0CG}} =
\left( {\begin{array}{*{20}c}
    0\\
    R \\
    0\\
 \end{array} } \right)
 \quad
  \underline{x}_{I,\overrightarrow{0FA}} =
\left( {\begin{array}{*{20}c}
    -l_{FA} cos \beta\\
    l_{FA} sin \beta\\
    0\\
 \end{array} } \right) 
 \quad
   \underline{x}_{I,\overrightarrow{0RA}} =
\left( {\begin{array}{*{20}c}
    l_{RA} cos \beta\\
    - l_{RA} sin \beta\\
    0\\
 \end{array} } \right)
 \end{split}
\end{equation*}
\begin{equation*}
\begin{split}
 \underline{x}_{B,\overrightarrow{0CG}} =
\left( {\begin{array}{*{20}c}
    -R sin \beta\\
    R cos \beta\\
    0\\
 \end{array} } \right)
 \quad
  \underline{x}_{B,\overrightarrow{0FA}} =
\left( {\begin{array}{*{20}c}
    -l_{FA}\\
    0\\
    0\\
 \end{array} } \right) 
 \quad
   \underline{x}_{B,\overrightarrow{0RA}} =
\left( {\begin{array}{*{20}c}
    l_{RA}\\
    0\\
    0\\
 \end{array} } \right)
 \end{split}
\end{equation*}

\noindent\underline{Velocity \& acceleration of CG}\\
For the absolute velocity and acceleration of the CG one finds
\begin{equation}
\begin{split}
 \underline{\dot{x}}_{I,\overrightarrow{0CG}} = 
\frac{d \underline{\dot{x}}_{I,\overrightarrow{0CG}}}{dt}+\left(\underline{\dot{\alpha}} + \underline{\dot{\beta}}\right) \times \underline{x}_{B,\overrightarrow{0CG}}= \left( {\begin{array}{*{20}c}
    -R  ~\left(\dot{\alpha} + \dot{\beta} \right)\\
    \dot{R}\\
    0\\
 \end{array} } \right)\\
 \end{split}\label{eq:vel_SP_I}
 \end{equation}
\begin{equation}
\begin{split}
\underline{\dot{x}}_{B,\overrightarrow{0CG}} = \underline{\underline{A}}_{B,I} ~ \underline{\dot{x}}_{I,\overrightarrow{0CG}}  =\left( {\begin{array}{*{20}c}
    -R  \left(\dot{\alpha} + \dot{\beta} \right) cos \beta - \dot{R} ~ sin \beta\\
    \dot{R}  ~cos \beta - R  \left(\dot{\alpha} + \dot{\beta} \right)  sin \beta \\
    0\\
 \end{array} } \right)\\
 \end{split}\label{eq:vel_SP_B}
 \end{equation}

\begin{equation}
 \begin{split}
  \underline{\ddot{x}}_{I,\overrightarrow{0CG}} =\frac{d \underline{\dot{x}}_{I,\overrightarrow{0CG}}}{dt}+\left(\underline{\dot{\alpha}} + \underline{\dot{\beta}}\right) \times \underline{\dot{x}}_{I,\overrightarrow{0CG}}=\\
  \left( {\begin{array}{*{20}c}
    - 2  \dot{R}  \left(\dot{\alpha} + \dot{\beta} \right) - R \left(\ddot{\alpha} + \ddot{\beta} \right)\\
    \ddot{R} - R \left(\dot{\alpha} + \dot{\beta} \right)^2 \\
    0\\
 \end{array} } \right)
 \end{split}\label{eq:Acc_SP}
 \end{equation}

\begin{equation}
 \begin{split}
  \underline{\ddot{x}}_{B,\overrightarrow{0CG}} = \underline{\underline{A}}_{B,I} ~ \underline{\ddot{x}}_{I,\overrightarrow{0CG}}=\\
  \left( {\begin{array}{*{20}c}
    - cos \beta \left[2  \dot{R}  \left(\dot{\alpha} + \dot{\beta} \right) + R \left(\ddot{\alpha} + \ddot{\beta} \right)\right] - sin \beta \left[ \ddot{R} - R \left(\dot{\alpha} + \dot{\beta} \right)^2\right] \\
    - sin \beta \left[2  \dot{R}  \left(\dot{\alpha} + \dot{\beta} \right) + R \left(\ddot{\alpha} + \ddot{\beta} \right)\right]  + cos \beta \left[ \ddot{R} - R \left(\dot{\alpha} + \dot{\beta} \right)^2\right]\\
    0\\
 \end{array} } \right)
 \end{split}\label{eq:Acc_SP_B}
 \end{equation}
As stated in Section \ref{sec:physical_model}, the accelerating vehicle follows a curve of constant $R$ with velocity $V(t)$; thus, $\ddot{\alpha} \neq 0$. Moreover, stationary steering is assumed, i.e. $\dot{\beta}=0$. Thus 
\begin{equation}
R = \frac{V}{\dot{\beta} + \dot{\alpha}} = \frac{V}{\dot{\alpha}} 
\label{eq:R-alpha-V}
\end{equation}
Hence, Eq. \eqref{eq:vel_SP_I} to \eqref{eq:Acc_SP_B} become

\begin{equation}
\begin{split}
 \underline{\dot{x}}_{I,\overrightarrow{0CG}} =  \left( {\begin{array}{*{20}c}
    -R  ~\dot{\alpha} \\
    0\\
    0\\
 \end{array} } \right)
 = \left( {\begin{array}{*{20}c}
    -V \\
    0\\
    0\\
 \end{array} } \right)\\
 \end{split}\label{eq:vel_SP_I_simple}
 \end{equation}
\begin{equation}
\begin{split}
\underline{\dot{x}}_{B,\overrightarrow{0CG}} =\left( {\begin{array}{*{20}c}
    -R \dot{\alpha}~ cos \beta \\
     - R  ~\dot{\alpha} ~ sin \beta \\
    0\\
 \end{array} } \right)=\left( {\begin{array}{*{20}c}
    -V~ cos \beta \\
     - V ~ sin \beta \\
    0\\
 \end{array} } \right)
 \end{split}\label{eq:vel_SP_B_simple}
 \end{equation}

\begin{equation}
\begin{split}
\underline{\ddot{x}}_{I,\overrightarrow{0CG}} =
  \left( {\begin{array}{*{20}c}
    - R \ddot{\alpha} \\
    - R \dot{\alpha}^2 \\
    0\\
 \end{array} } \right)
 \end{split}\label{eq:Acc_SP_Simple}
 \end{equation}

\begin{equation}
 \begin{split}
  \underline{\ddot{x}}_{B,\overrightarrow{0CG}} =
  \left( {\begin{array}{*{20}c}
    - cos \beta \left[ R \ddot{\alpha} \right] + sin \beta \left[  R \dot{\alpha}^2\right] \\
    - sin \beta \left[ R \ddot{\alpha}\right]  - cos \beta \left[  R \dot{\alpha}^2\right]\\
    0\\
 \end{array} } \right)
 \end{split}\label{eq:Acc_SP_B_Simple}
 \end{equation}
\underline{Conservation of momentum in body coordinate system in CG}\\
For this system the equations for conservation of momentum and angular momentum about the center of gravity are:

\begin{equation}
m_{veh}  \underline{\ddot{x}}_{B,\overrightarrow{0CG}} =\sum_{i}^{} F_{i,B}
\end{equation}

\begin{equation}
\begin{split}
m_{veh}\left( {\begin{array}{*{20}c}
    - cos \beta \left[ R \ddot{\alpha} \right] + sin \beta \left[  R \dot{\alpha}^2\right] \\
    - sin \beta \left[ R \ddot{\alpha}\right]  - cos \beta \left[  R \dot{\alpha}^2\right]\\
    0\\
 \end{array} } \right) &=
   \left( {\begin{array}{*{20}c}
    0\\
    0\\
		-F_{N}\\
\end{array} } \right) +\left( {\begin{array}{*{20}c}
    -F_{\parallel, RA}\\
    F_{\perp, RA}\\
		F_{N,RA}\\
\end{array} } \right) +
\left( {\begin{array}{*{20}c}
    cos \delta & - sin \delta & 0 \\
    sin \delta &   cos \delta & 0\\
		0 & 0 & 1 \\
\end{array} } \right)
\left( {\begin{array}{*{20}c}
    -F_{\parallel, FA}\\
    F_{\perp, FA}\\
		F_{N,FA} \\  
\end{array} } \right)=\\
&=\left( {\begin{array}{*{20}c}
    -F_{\parallel, RA}-F_{\parallel, FA} cos \delta - F_{\perp, FA} sin \delta\\
    F_{\perp, RA} - F_{\parallel, FA} sin \delta + F_{\perp, FA} cos \delta\\
		-F_{N}+F_{N,RA}+F_{N,FA}
\end{array} } \right)
\end{split}\label{eq:Cons_Momentum}
\end{equation}

\begin{equation}
\begin{split}
   \frac{d \underline{L}_{CG, B}}{dt} =  \underline{l}_{FA, B} \times \underline{F}_{FA, B}+\underline{F}_{RA, B} \times \underline{l}_{RA, B} =\\
        \left( {\begin{array}{*{20}c}
        -F_{\parallel, FA} cos \delta - F_{\perp, FA} sin \delta\\
        - F_{\parallel, FA} sin \delta + F_{\perp, FA} cos \delta\\
        F_{N,FA}\\
        \end{array} } \right) \times
        \left( {\begin{array}{*{20}c}
        -l_{FA}\\
        0\\
        0\\
        \end{array} } \right)+				
				\left( {\begin{array}{*{20}c}
        - F_{\parallel, RA}\\
        F_{\perp, RA}\\
        F_{N,RA}\\
        \end{array} } \right) \times
        \left( {\begin{array}{*{20}c}
        l_{RA}\\
        0\\
        0\\
        \end{array} } \right)
         =\\ 
        \left(0;   F_{N,RA}l_{RA}-F_{N,FA}l_{FA}; -F_{\perp, RA} l_{RA} + (-F_{\parallel, FA} sin \delta + F_{\perp, FA} cos \delta)l_{FA}\right)^T
\end{split}\label{eq:Cons_Ang_Momentum}
\end{equation}
With 
\begin{equation}
    \frac{d \underline{L}_{CG}}{dt}=\left(0; 0;\ddot{\alpha}~m_{veh}~0\right)^T = 0
\end{equation}
From Eqs. \eqref{eq:Cons_Ang_Momentum}, \eqref{eq:Cons_Momentum} one finds that
\begin{equation}
F_{\perp, FA}=\frac{F_{\parallel, FA} R \left(l_{FA} sin \delta cos \beta - l_{RA} sin \left(\beta - \delta \right)\right) - F_{\parallel, RA} R l_{RA} sin \beta - V^2 l_{RA} m_{veh}  }{R \left( l_{FA} cos \beta cos \delta + l_{RA} cos\left(\beta - \delta \right)\right)},
\label{eq:F_FA_perp_final}
\end{equation} 
\begin{equation}
    F_{\perp, RA}=\frac{- l_{FA}\left( F_{\parallel, FA} R sin \beta + F_{\parallel, RA} R sin \beta cos \delta + V^2 m_{veh} cos \delta\right)}{R \left( l_{FA} cos \beta cos \delta + l_{RA} cos\left(\beta - \delta \right)\right)} .
		\label{eq:F_RA_perp_final}
\end{equation}
The second entry of Eq. \eqref{eq:Cons_Ang_Momentum} and the third entry of Eq. \eqref{eq:Cons_Momentum} yield that
\begin{subequations} 
\begin{equation}
F_{N,FA}= \frac{l_{RA}}{l}F_{N},
\label{eq:Normal-Force-FA}
\end{equation}
\begin{equation}
F_{N,RA}= \frac{l_{FA}}{l}F_{N}.
\label{eq:Normal-Force-RA}
\end{equation}
\end{subequations}

The friction force correlates to the normal force according to $F_{\mu}= \mu ~ F_N$. $\mu=\mu(\mu^{\star},w_{T},h_{T})$ 
is the friction coefficient that depends on the road-tire friction coefficient $\mu^{\star}$, the tire width $w_{T}$ and the height of the tire wall $h_{T}$. Within the scope of this work, we assume $\mu = \mu^{\star}$ and neglect the influence of $w_{T}$ and $h_{T}$.
For each axle, $T_{X}=F_{\parallel, X}~ r_{dyn}$ with $X \in \{FA; RA\}$ holds. If a vehicle has a single engine and a differential gear to distribute the torque $T_{mot}$ to the front and rear axle,  the total transmission ratios for the front and rear axle are:
\begin{subequations}
\begin{equation}
i_{tot, FA}= i_{gear}~i_{Diff, FA}=\frac{n_{mot}}{n} i_{Diff, FA}= \frac{T_{FA}}{T_{mot}},
\label{eq:i_tot_FA}
\end{equation} 
\begin{equation}
i_{tot, RA}= i_{gear}~i_{Diff, FA}=\frac{n_{mot}}{n} i_{Diff, RA}= \frac{T_{RA}}{T_{mot}},
\label{eq:i_tot_RA}
\end{equation}
\end{subequations}
respectively. Thus, the longitudinal forces are   
\begin{subequations}
\begin{equation} 
    F_{\parallel, FA}=\frac{T_{mot} ~ i_{tot, FA}}{r_{dyn}},
\end{equation} \label{eq:Accelerating_Force_FA}
\begin{equation} 
    F_{\parallel, RA}=\frac{T_{mot} ~ i_{tot, RA}}{r_{dyn}},
\end{equation} \label{eq:Accelerating_Force_RA}
\end{subequations}
respectively.
If dedicated engines for the front and rear axle are used instead,
\begin{subequations}
\begin{equation} 
    F_{\parallel, FA}=\frac{T_{mot, FA}~i_{FA}}{r_{dyn}},
\end{equation} \label{eq:Accelerating_Force_FA2}
\begin{equation} 
    F_{\parallel, RA}=\frac{T_{mot, RA}~i_{RA}}{r_{dyn}},
\end{equation} \label{eq:Accelerating_Force_RA2}
\end{subequations}
respectively.

 \bibliographystyle{elsarticle-num-names}

\addcontentsline{toc}{chapter}{Bibliography}
\end{document}